# Amplitude of vortical turbulence in crossed fields in mass-separator at optimum parameters

*V.I.Maslov, I.P.Levchuk, I.N.Onishchenko, V.B.Yuferov*

*NSC Kharkov Institute of Physics & Technology, 61108 Kharkov, Ukraine*

*vmaslov@kipt.kharkov.ua*

Properties and excitation of the vortical turbulence, excited in a cylindrical radially inhomogeneous plasma in crossed radial electric and longitudinal magnetic fields, are considered. The expression for the vortex amplitude of saturation has been derived.

## 1. INTRODUCTION

It is well known from numerous numerical simulations (see, for example, [1]) and from experiments (see, for example [2]) that electron density nonuniformity in kind of discrete vortices are long-living structures. In experiments [2] a rapid re-organization of discrete electron density nonuniformity has been observed in the spatial distribution of vorticity in pure electron plasma when a discrete vortex has been immersed in an extended distribution of the background vorticity. In plasma lens [3-11] for high-current ion beam focusing, in mass-separator [12-16], in nuclear fusion [17-19] and in plasma-optical device for the elimination of droplets in cathodic ARC plasma coating [20], a vortical turbulence has been excited in crossed radial electrical and longitudinal magnetic fields [21-23] by gradient of external magnetic field or by gradient of plasma density. This turbulence is a distributed vorticity. In this paper the amplitude of the vortex saturation in cylindrical radially inhomogeneous plasma in crossed radial electric and axial magnetic fields in the mass-separator [14] for the optimal parameters are investigated theoretically. It is shown that as the parameters tend to the optimal ones the amplitude of excited vortices tends to zero. This allows to specify a range of parameters of the experimental setup for which the vortical turbulence is suppressed.

Excitation of the vortices in the approximation of strongly magnetized electrons and weakly magnetized ions is studied analytically. Plasma is distributed inhomogeneously in radial direction and is a cylinder of finite length, placed in a magnetic field of short coil.

It has been shown in [6] that the system is unstable relatively to excitation of oscillating fields in crossed fields. Excitation of the oscillation is realized as a result of a positive radial gradient of magnetic field of short coil and negative radial gradient of the plasma density.

The ions drift on angle, $\theta$, with a velocity $V_{\theta oi} = -eE_{or}/m_i\omega_{Hi}$, $\omega_{Hi} = eH_0/m_ic$, because the crossed configuration of radial electric, $E_{or}$, and longitudinal magnetic fields, $H_0$, is maintained in the separator.

The perturbation of the plasma particle density leads to appearance of an electric field in the vicinity of the perturbation. Therefore, the crossed configuration of fields is realized in the vicinity of the perturbation. Thus, the dynamics of plasma particles is vortical in the field of perturbation.

In this paper, the properties and the excitation of vortical perturbations in crossed fields in the mass-separator are studied theoretically.

## 2. DERIVATION OF DISPERSION RELATION

We use the hydrodynamic equations for plasma electrons and ions and the Poisson equation

$$\partial_t \vec{V}_e + (\vec{V}_e \vec{\nabla}) \vec{V}_e = \frac{e}{m_e} \vec{\nabla}\varphi + [\vec{\omega}_{He}, \vec{V}_e] - \frac{V_{the}^2}{n_e} \vec{\nabla} n_e,$$

$$\partial_t n_e + \vec{\nabla}(n_e \vec{V}_e) = 0, \quad (1)$$

$$\partial_t \vec{V}_i + (\vec{V}_i \vec{\nabla}) \vec{V}_i = -\frac{e}{m_i} \vec{\nabla}\varphi - [\vec{\omega}_{Hi}, \vec{V}_i] - \frac{V_{thi}^2}{n_i} \vec{\nabla} n_i,$$

$$\partial_t n_i + \vec{\nabla}(n_i \vec{V}_i) = 0, \quad \vec{\nabla}\varphi \equiv \vec{\nabla}\phi - \vec{E}_{r0},$$

$$\Delta\varphi = 4\pi e(\Delta n + \delta n_v), \quad \Delta n + \delta n_v = n_e - n_i. \quad (2)$$

Quasistationary $\Delta n$ determines the value of the radial electric field $E_{r0}$, $\delta n_v$ is the perturbation of the plasma density in the vortex. From (1), (2) one can derive the approximate equations

$$d_t\left(\frac{\alpha - \omega_{He}}{n_e}\right) = 0, \quad d_t = \partial_t + \vec{V}_{e\perp}\vec{\nabla}_\perp, \quad \alpha \equiv \vec{e}_z \text{rot}\vec{V}_e \quad (3)$$

$$d_t\left(\frac{\alpha_i - \omega_{Hi}}{n_i}\right) = 0, \quad d_{ti} = \partial_t + \vec{V}_{i\perp}\vec{\nabla}_\perp, \quad \alpha_i \equiv \vec{e}_z \text{rot}\vec{V}_i$$

From (1), (2) one can derive in linear approximation

$$\vec{V}_e = \frac{e}{m_e\omega_{He}}[\vec{e}_z, \vec{\nabla}\varphi], \quad \vec{V}_i = \frac{e}{m_i\omega_{Hi}}[\vec{e}_z, \vec{\nabla}\varphi] \quad (4)$$

$$\alpha = \frac{2eE_{r0}}{rm_e\omega_{He}} + \frac{e}{m_e\omega_{He}}\Delta\phi, \quad \alpha_i = \frac{2eE_{r0}}{rm_i\omega_{Hi}} + \frac{e}{m_i\omega_{Hi}}\Delta\phi \quad (5)$$

From (2), (5) it approximately follows that vortical motion begins (see Fig. 1) as soon as the perturbation $\delta n$ appears.

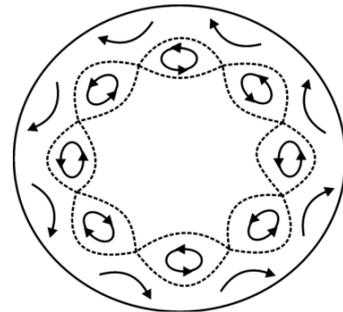

Fig. 1. The structure of electron trajectories in the train of vortices in a system that rotates with $\omega_{ph} = V_{ph}/r_v$

We consider magnetized ions, i.e., we expect that the following inequality holds

$$R \geq eE_{r0}/m_i\omega_{ci}^2 . \quad (6)$$

R is the radius of the system, $\omega_{He}$, $\omega_{Hi}$ are the cyclotron frequencies of the plasma electrons and ions, $n_e$, $n_i$ are the density of plasma electrons and ions. Previously authors have shown [6] that the dynamics of electrons in crossed radial electric and axial magnetic fields in the approximation of a homogeneous system in the longitudinal direction is described approximately according to (3) by the following equation

$$d_t\left(\frac{\omega_{He}}{n_e}\right) = 0 . \quad (7)$$

In linear approximation one can derive

$$\vec{V}_e = \vec{V}_{\theta 0} + \left(\frac{e}{m_e\omega_{He}}\right)\left[\vec{e}_z, \vec{\nabla}_\perp \phi\right] \quad (8)$$

$\vec{V}_{\theta 0}$ is the drift velocity along the azimuth of the plasma electrons in crossed fields; $\phi$ is the electric potential of the vortical perturbation. Also in the linear approximation from (7) we obtain

$$-\frac{\omega_{He}}{n_0^2}\left(\partial_t + \omega_{\theta o}\partial_\theta\right)\delta n_e + \delta V_r \partial_r\left(\frac{\omega_{He}}{n_0}\right) = 0 . \quad (9)$$

$n_e = n_o + \delta n_e$, $\omega_{\theta o} \equiv V_{\theta o}/r$. From (8), (9) we have

$$\delta n_e = -\frac{k_\theta}{(\omega - V_{\theta o}k_\theta)}\phi\partial_r\left(\frac{cn_0}{H_0}\right) . \quad (10)$$

The same expression can be obtained for $\delta n_i$

$$\delta n_i = \frac{k_\theta}{(\omega - V_{\theta oi}k_\theta)}\varphi\partial_r\left(\frac{cn_0}{H_0}\right) . \quad (11)$$

$\vec{V}_{\theta 0i}$ is the drift velocity along the azimuth of the plasma ions in crossed fields.

Substituting (10), (11) in the Poisson equation, we derive the dispersion relation in the case of magnetized ions

$$1 - \frac{k_\theta\partial_r\left(\omega_{pi}^2/\omega_{Hi}\right)}{(\omega - k_\theta V_{\theta oi})k^2} - \frac{k_\theta\partial_r\left(\omega_{pe}^2/\omega_{He}\right)}{(\omega - k_\theta V_{\theta o})k^2} = 0 . \quad (12)$$

$k$ is the wave vector, $k_\theta$ is the azimuth wave vector.

## 3. INSTABILITY SUPPRESSION IN CROSSED FIELDS IN MASS-SEPARATOR AT OPTIMUM PARAMETERS

Development of plasma instability continues as long as distributed plasma particles are grouped (or compressed). The instability is suppressed when the plasma particle compressing stops. And if the perturbation of the ion density $\delta n_i$ equals zero, then according to the dispersion relation the growth rate of instability development also equals zero. The maximum amplitude of the vortices, $\phi_{vm}$, is determined by the condition that the magnetic force is no longer keeps the particles of the vortex, rotating around its axis along a closed trajectories, and the particles can be extended across the magnetic field. Thus particle compressing stops. Thus, from the condition of an imbalance of forces that provide movement along a closed trajectories of the particles

$$\frac{m_i V_{\theta i}^2}{r} + eE_r \geq m_i\omega_{Hi}V_{\theta i} . \quad (13)$$

one can find the vortex saturation amplitude, $\phi_{sm}$. From (13) it follows that the particles of the vortex when the inequality is correct

$$\frac{\omega_{pi}^2}{n_0}(\Delta n + \delta n_v) \geq \frac{\omega_{Hi}^2}{2} \quad (14)$$

can be extended across the magnetic field. $E_r$ is determined by an external electric field $E_{r0} = -4\pi e\Delta n$ and by the electric field of vortex, the plasma density perturbation in which equals $\delta n_v = \delta n_e - \delta n_i$. Then from the Poisson equation we have

$$\Delta\phi = 4\pi e\delta n_v , \quad \phi \approx -4\pi(e/k^2)\delta n_v . \quad (15)$$

We consider the case of a strongly magnetized electrons and of a weakly magnetized ions. Using (14), (15), we find that the amplitude of the vortex is stabilized at

$$\phi_{vm} \approx \frac{m_i}{ek^2}\left[\frac{\omega_{Hi}^2}{2} - \omega_{pi}^2\frac{\Delta n}{n_0}\right] \quad (16)$$

From (16) one can see that if $\Delta n$ is close to

$$\Delta n \approx \frac{H_0^2}{8\pi m_i c^2} , \quad (17)$$

the vortical perturbations are suppressed. From (17) it follows

$$H_0 \propto \sqrt{m_i} .$$

Thus, the instability is suppressed in the case of strongly magnetized ions in a collisionless plasma, when there is no the relative drift of the plasma electrons and ions. Also, the instability is suppressed at the optimum magnetic field (17).

## 4. CONCLUSION

Properties and excitation of the vortical turbulence, excited in a cylindrical radially inhomogeneous plasma in crossed radial electric and longitudinal magnetic fields, have been described.